\documentclass[nohyper,12pt]{JHEP3}
\usepackage{epsfig}
\usepackage{amsfonts,amssymb}

\newcommand{\be}{\begin{equation}}
\newcommand{\ee}{\end{equation}}
\newcommand{\bea}{\begin{eqnarray}}
\newcommand{\eea}{\end{eqnarray}}
\newcommand{\ba}{\begin{array}}
\newcommand{\ea}{\end{array}}

\def\bbox{{\,\lower0.9pt\vbox{\hrule \hbox{\vrule height 0.2 cm
\hskip 0.2 cm \vrule height 0.2 cm}\hrule}\,}}
\newcommand{\dsl}{\pa \kern-0.5em /}

\newcommand{\EQ}{\begin{equation}}
\newcommand{\EN}{\end{equation}}

\def\bbox{{\,\lower0.9pt\vbox{\hrule \hbox{\vrule height 0.2 cm
\hskip 0.2 cm \vrule height 0.2 cm}\hrule}\,}}

\newcommand{\pa}{\partial}

\newcommand{\la}{\lambda}

\def\be{\begin{equation}}
\def\ee{\end{equation}}
\def\ba{\begin{eqnarray}}
\def\ea{\end{eqnarray}}
\def\la{~\mbox{\raisebox{-.6ex}{$\stackrel{<}{\sim}$}}~}

\def\bq{\begin{quote}}
\def\eq{\end{quote}}

\makeatletter
\def\vereq#1#2{\lower3pt\vbox{\baselineskip1.5pt \lineskip1.5pt
\ialign{$\m@th#1\hfill##\hfil$\crcr#2\crcr\sim\crcr}}}
\makeatother  at 10truept

\newcommand{\beq}{\begin{equation}}
\newcommand{\eeq}{\end{equation}}
\newcommand{\beqa}{\begin{eqnarray}}
\newcommand{\eeqa}{\end{eqnarray}}

\def\la{~\mbox{\raisebox{-.6ex}{$\stackrel{<}{\sim}$}}~}

\def\ltap{\ \raise.3ex\hbox{$<$\kern-.75em\lower1ex\hbox{$\sim$}}\ }
\def\gtap{\ \raise.3ex\hbox{$>$\kern-.75em\hbox{$\sim$}}\ }
\def\gl{\ \raise.5ex\hbox{$>$}\kern-.8em\lower.5ex\hbox{$<$}\ }
\def\roughly#1{\raise.3ex\hbox{$#1$\kern-.75em\lower1ex\hbox{$\sim$}}}

\title{Two gravitational shock waves on the $AdS_3$ brane}

\author{Mohamed Anber and Lorenzo Sorbo\\
Department of Physics, University of Massachusetts, Amherst, MA 01003, USA \\
E-mail: \email{{\tt manber@physics.umass.edu}}, \email{{\tt sorbo@physics.umass.edu}}}

\abstract{A gravitational shock wave is a solution to Einstein equations describing the gravitational field of a massless particle. We obtain such a geometry for a particle moving on a $AdS_3$ brane embedded in a $AdS_4$ bulk (the lower dimensional version of the ``locally localized gravity'' model of Karch and Randall). In order to do this, we use two different techniques. First, we solve directly Einstein equations sourced by a massless particle. Then we boost to the speed of light the $AdS_3$ brane black hole solution of Emparan et al.~\cite{Emparan:1999fd} while sending its mass parameter to zero. Surprisingly, we obtain two different results. We discuss the origin of these two different solutions both in the bulk and in the CFT picture. As a by-product, we derive the expression for the shock wave associated to a transversally accelerating particle in $AdS_4$.}

\keywords{Localization of Gravity, AdS/CFT, Braneworlds}

\begin{document} 

\section{Introduction}

The Randall-Sundrum model~\cite{Randall:1999ee,Randall:1999vf} offers a new way, alternative to the Kaluza-Klein compactification, to achieve lower dimensional gravity at large distances. The model is realized by cutting Anti-de Sitter ($AdS$) space with a codimension-one brane, where gravity is localized. Depending on its tension, the brane can have the geometry of Minkowski, de Sitter ($dS$) or Anti-de Sitter space. Of course, the phenomenologically most interesting case is that of a minkowskian 3-brane. However, other configurations have a profound theoretical interest. In the present paper we will be concerned with $AdS$ branes, and we will focus in particular on the case of a $2$-brane in a $3+1$ dimensional bulk.

The mechanism of gravity localization works in very different ways depending on the brane tension. For a supercritical tension (leading to a $dS$ brane), the brane is accelerating in the $AdS$ bulk, so that the $AdS$ boundary is hidden behind a Rindler horizon. A brane observer sees a spectrum of Kaluza-Klein gravitons that contains a zero mode and a continuum of massive modes separated from the zero mode by a mass gap. If we decrease the brane tension, the gap reduces and the spectrum eventually collapses, for a Minkowski brane, to a continuum that still starts from a zero mode. If we keep lowering the tension we obtain an $AdS$ brane, and the brane observer can now see the boundary of the bulk $AdS$ space and its infinite volume. This implies that the zero mode of the graviton is not normalizable and - strictly speaking - we do not expect to see localized gravity any more. By continuity, however, if the brane $AdS$ radius is large enough gravity should still appear localized at least in some regime. This is precisely what happens in the {\em locally localized gravity} model of Karch and Randall~\cite{Karch:2000ct}. In this case brane gravity looks like usual massless gravity from distances of the order of the bulk $AdS$ radius all the way to distances much larger than the brane $AdS$ radius. At much larger scales, however, some deviation emerges, showing that the graviton has actually an (ultrasmall) mass. For this reason, the Karch-Randall is especially interesting: its low energy regime describes a consistent model of massive gravity that starts with a generally covariant action. 

\vskip0.5cm

The first exact solution for a localized source in the context of the Karch-Randall model has been presented in~\cite{Kaloper:2005wq}. This solution is a gravitational {\em shock wave}, the gravitational field of a massless particle. The solution of~\cite{Kaloper:2005wq} allowed to derive a number of properties of the locally localized gravitational field, including the couplings to matter of the Kaluza-Klein modes of the graviton. 

In general a shock wave solution can be found in two different ways. A first possibility is to solve directly Einstein's equations for a null source. In this case, the ``cut and paste'' trick of Dray and 't~Hooft \cite{Dray:1984ha} is especially useful. The second option is to start from the gravitational field of a massive particle at rest, and then boost the particle to the speed of light while sending its mass to zero, so that the momentum of the particle remains finite. This technique was used in the original work by Aichelburg and Sexl~\cite{Aichelburg:1970dh}. 

In the case of a $AdS_4$ 3-brane studied in~\cite{Kaloper:2005wq} only the first technique could be used, since no solution for a massive particle on the $AdS_4$ brane is known\footnote{The only such solution explicitly known in this case is the bulk black string discussed e.g. in~\cite{Chamblin:2004vr}, that however corresponds to a source that extends through the whole bulk.}. In the lower dimensional case of a $AdS_3$ brane, however,  solutions associated to brane localized matter were found by Emparan et al.~\cite{Emparan:1999fd,Emparan:1999wa}. These solutions are obtained by cutting the $AdS$ C-metric (that describes a particle in $AdS_4$ space attached to a string that accelerates it~\cite{Plebanski:1976gy}) with an $AdS_3$ brane.

For this reason, in this paper we will study shock waves in the lower dimensional case of a $AdS_3$ 2-brane embedded in a $AdS_4$ bulk. This will allow us to use both
of the techniques described above. In section 3 we will derive the shock wave on the $AdS_3$ brane starting directly from a relativistic source and using Dray and 't~Hooft's technique. Then, in section 4, we derive the shock wave solution by boosting the $AdS_3$ brane black hole of Emparan et al.~\cite{Emparan:1999fd}. In order to derive this result, we compute the metric that corresponds to boosting to the speed of light the $AdS$ C-metric in a direction orthogonal to the string. To our knowledge, this is a new solution to Einstein equations, not present in the existing literature. 

Surprisingly enough, the two solutions of sections 3 and 4 {\em do not agree} with each other. This is our main result, and we will discuss the origin of this discrepancy in section 5.

\vskip0.5cm

One of the most interesting aspects of the Randall-Sundrum model is that it is conjectured to enjoy a  dual interpretation in terms of a conformal field theory (CFT) coupled to gravity. According to this duality, the classical dynamics in the $AdS_{d+1}$ bulk cut by a brane describes the quantum dynamics of the dual CFT, in the planar 
limit of a large $N$ expansion, coupled to classical  gravity in $d$ dimensions. This implies that solutions of {\em classical} $d+1$ dimensional Einstein equations can be mapped into solutions of  $d$-dimensional Einstein equations that include a  CFT stress-energy tensor considered at the {\em quantum} level in the planar limit.

The conjectured duality has passed several tests, such as those of~\cite{Gubser:1999vj,Duff:2000mt}. Some of these tests concern the case of a $AdS$ brane: in a series of papers~\cite{Porrati:2001gx,Porrati:2001db,Porrati:2003sa,Duff:2004wh} the generation of an ultralight mass for the graviton has been explained in the CFT picture as a quantum effect of the coupling of the CFT to gravity, similarly to what happens in 1+1 dimensional Quantum Electrodynamics where a mass for the photon is generated by quantum effects~\cite{Schwinger:1962tp}. 

In section 5 we will discuss the CFT interpretation of the different solutions found in sections 3 and 4. As we will see, the origin of these different solutions can be traced back to the fact that, for a $AdS$ brane, the CFT on the brane does not encode all of the bulk degrees of freedom, but only part of them. The remaining part of bulk degrees of freedom can  be mapped onto a CFT that lives on the boundary of the bulk $AdS$ space. As we will  see, the two different solutions of section 3 and 4 can be argued to emerge from the fact that the CFT at the boundary of $AdS_4$ are in different states.

\section{Shock wave on $AdS_{3}$}

Let us start by reviewing the shock wave geometry in  $AdS_3$. This metric has been found by Sfetsos~\cite{Sfetsos:1994xa} by solving directly Einstein equations for a null source. More recently Cai and Griffiths~\cite{Cai:1999dz} have recovered this solution by boosting a conical singularity in $AdS_3$ while sending the mass to zero. Both methods give the same result, that can be represented as follows. We start from four dimensional space with coordinates $Z_0,\,Z_1,\,Z_2,\,Z_3$ and metric $ds^2=-dZ_0^2+dZ_1^2+dZ_2^2-dZ_3^2$. Empty $AdS_3$ (with $AdS$ radius $\ell_3$) is the hyperboloid $-Z_0^2+Z_1^2+Z_2^2-Z_3^2=-\ell_3^2$. It is possible to write the $AdS_3$ metric in light cone coordinates $u,\,v,\,\chi$ as follows
\begin{eqnarray}\label{ads3lightcone}
&&Z_0=\frac{v-u}{1-uv/\ell_3^2}\,\,,\nonumber\\
&&Z_1=\frac{u+v}{1-uv/\ell_3^2}\,\,,\nonumber\\
&&Z_2=\ell_3\,\frac{1+uv/\ell_3^2}{1-uv/\ell_3^2}\,\sinh{\chi}\,\,,\nonumber\\
\label{u v coordinates for ads3}
&&Z_3=\ell_3\,\frac{1+uv/\ell_3^2}{1-uv/\ell_3^2}\,\cosh{\chi}\,\,,
\end{eqnarray}
so that the metric is
\begin{equation}\label{emptyads3}
ds^2_{\mathrm {Empty\ AdS_3}}=\frac{4\,du\, dv}{\left(1-uv/\ell_3^2\right)^2}+\ell_3^2\,\left(\frac{1+uv/\ell_3^2}{1-uv/\ell_3^2}\right)^2\,d\chi^2\,\,.
\end{equation}

In terms of these coordinates, the shock wave associated to a particle with momentum $p$ moving along the null trajectory $u=0$ reads\footnote{Our convention in the definition of the $d$-dimensional Planck mass is that Einstein equations read $G_{\mu\nu}=T_{\mu\nu}/M_d^{d-2}$.}~\cite{Sfetsos:1994xa,Cai:1999dz}
\begin{equation}\label{shockads3}
ds^2_{\mathrm {Shock\ AdS_3}}=ds^2_{\mathrm {Empty\ AdS_3}}+2\,\frac{p\,\ell_3}{M_3}\,e^{-\left|\chi\right|}\,\delta\left(u\right)\,du^2\,\,.
\end{equation}
The shockwave in this case represents a deficit angle in the background space. Since gravity in three dimensions is not dynamical, this solution has topological nature.

\section{Shock wave on $AdS_{3}$ brane embedded in $AdS_4$ space}

In this section we use the Dray and 't Hooft technique~\cite{Dray:1984ha} to derive the metric describing a shock wave on a $AdS_3$ brane (with $AdS$ radius $\ell_3$) embedded in $AdS_4$ space (radius $\ell_4$).

Our starting point is  empty $AdS_4$, that we describe as the hyperboloid $-W_0^2+W_1^2+W_2^2+W_3^2-W_4^2=-\ell_4^2$ embedded in a five-dimensional space with metric $ds^2=-dW_0^2+dW_1^2+dW_2^2+dW_3^2-dW_4^2$. A convenient choice of coordinates is the following
\begin{eqnarray}\label{ads3inads4}
&&W_0=\frac{\ell_4/\ell_3}{\sin\left(z/\ell_3\right)}\,Z_0\,\,,\nonumber\\
&&W_1=\ell_4\,\frac{\cos\left(z/\ell_3\right)}{\sin\left(z/\ell_3\right)}\,\,,\nonumber\\
&&W_2=\frac{\ell_4/\ell_3}{\sin\left(z/\ell_3\right)}\,Z_1\,\,,\nonumber\\
&&W_3=\frac{\ell_4/\ell_3}{\sin\left(z/\ell_3\right)}\,Z_2\,\,,\nonumber\\
&&W_4=\frac{\ell_4/\ell_3}{\sin\left(z/\ell_3\right)}\,Z_3\,\,,
\end{eqnarray}
where $Z_0,\,...,\,Z_3$ are given in eq.~(\ref{ads3lightcone}). 

In this coordinate system $AdS_{4}$ is foliated into $AdS_{3}$ slices
\begin{equation}\label{ads4slicedads3}
ds^2_{\mathrm {AdS_4}}=\frac{\ell_{4}^{\,2}}{\ell_{3}^{\,2}\sin^2\left(z/\ell_{3}\right)}\left[ds^2_{\mathrm {AdS_{3}}}+dz^2 \right] \mbox{ ,}
\end{equation}
where $ds^2_{\mathrm {AdS_{3}}}$ is the same as in equation~(\ref{emptyads3}). Slices of constant $z$ cut $AdS_3$ spaces with radius $\ell=\ell_4/\sin\left(z/\ell_3\right)$. In particular, we are interested in a brane with curvature $\ell_3$ so that we place it at $z_0$, where $z_0$ is defined by
\begin{equation}
\sin\left(z_0/\ell_3\right)=\ell_4/\ell_3\,.
\end{equation}

The Karch-Randall construction requires the full metric to be $Z_2$-symmetric across the brane. This can be achieved by replacing $z\rightarrow \left|z\right|+z_0$ in the metric above, so that~(\ref{ads4slicedads3}) becomes~\cite{Kaloper:1999sm}
\begin{equation}\label{ads4slicedads3new}
ds^2_{\mathrm {KR}}=\frac{\ell_{4}^{\,2}}{\ell_{3}^{\,2}\sin^2\left[\left(\left|z\right|+z_0\right)/\ell_{3}\right]}\left[ds^2_{\mathrm {AdS_{3}}}+dz^2 \right] \mbox{ .}
\end{equation}

Now we introduce our source: the only nonvanishing component of the stress energy tensor of a massless particle with momentum $p$ moving along the line $\chi=0$, $u=0$, $z=0$,  reads
\begin{equation}
T_{uu}^{\mbox{\scriptsize{particle}}}=\frac{2\,p}{\ell_3}\delta(u)\delta(z)\delta(\chi)\mbox{ .}
\end{equation}
For such a source, and following Dray and 't~Hooft~\cite{Dray:1984ha}, we look for the shock wave metric in the form (see also the appendix)
\begin{equation}\label{shockkr}
ds^2_{\mathrm {Shock\ KR}}=ds^2_{\mathrm {KR}}-\frac{4\,\ell_4^2}{\ell^2_3\,\sin^2\left[\left(\left|z\right|+z_0\right)/\ell_3\right]} f\left(z,\,\chi\right)\,\delta\left(u\right)\,du^2\,\,,
\end{equation}
where the equation for $f\left(z,\,\chi\right)$ reads
\begin{equation}\label{eqf}
\partial_{z}^{2} f-\frac{2}{\ell_{3}}\cot\left[\left(\left|z\right|+z_{0}\right)/\ell_{3}\right]\partial_{\left|z\right|}f+\frac{1}{\ell_3^2}\left(\frac{\partial^2}{\partial \chi^2}-1\right)f=\frac{p}{M_{4}^2\ell_{3}}\delta(\chi)\delta(z)\mbox{ .}
\end{equation}
Defining the variable $\zeta\equiv z/\ell_{3}$, and decomposing $f(\zeta,\,\chi)=\int_{-\infty}^{\infty} dq\,e^{iq\chi}\,\psi_{q}(\zeta)$ we get
\begin{equation}
\label{psiequationads}
\partial_{\zeta}^2\psi_{q}-2\cot\left(\left| \zeta\right| +\zeta_0 \right)\partial_{\left| \zeta\right|}\psi_{q}-(1+q^2)\psi_{q}=\frac{p}{2\pi M_{4}^2}\delta(\zeta) \mbox{ .}
\end{equation}
The solution of this equation~\cite{Stegun,bateman} is a linear combination of 
\begin{equation}
\cos\left(\left|\zeta\right|+\zeta_{0}\right)\,\sinh\left[q\left(\left|\zeta\right|+\zeta_{0}\right)\right]-q\sin\left(\left|\zeta\right|+\zeta_{0}\right)\,\cosh\left[(q\left(\left|\zeta\right|+\zeta_{0}\right)\right]{\phantom .}
\end{equation}
and
\begin{equation}
\cos\left(\left|\zeta\right|+\zeta_{0}\right)\,\cosh\left[q\left(\left|\zeta\right|+\zeta_{0}\right)\right]-q\sin\left(\left|\zeta\right|+\zeta_{0}\right)\,\sinh\left[(q\left(\left|\zeta\right|+\zeta_{0}\right)\right].
\end{equation}

Next, {\em we require our solution to be regular at the $AdS_{4}$ boundary}, and hence we impose $\psi_{q}(\left| \zeta\right|+\zeta_{0}=\pi)=0$ to get 
\begin{eqnarray}
\psi_q \left(\zeta\right)=N_q\left\{\cos\left(\left|\zeta\right|+\zeta_{0}\right)\,\sinh\left[ q\left(\left|\zeta\right|+\zeta_0-\pi\right)\right] -
q\sin\left(\left|\zeta\right|+\zeta_{0}\right)\,\cosh\left[q(\left|\zeta\right|+\zeta_{0}-\pi) \right] \right\}.\nonumber\\
\end{eqnarray}
The normalization constant $N_q$ can be found by integrating (\ref{psiequationads}) on a small interval around $\zeta=0$, so that the final expression for the shock wave reads
\begin{eqnarray}\label{fkr}
f\left(\zeta,\,\chi\right)&&=-\frac{p\,\ell_3}{2\pi\,M_4^2\,\ell_4}\,\int_0^{+\infty} dq \,\frac{\cos\, q\chi}{\left(1+q^2\right)\,\sinh\left[q\left(\zeta_0-\pi\right)\right]}\times\\
&&\times\left\{ \cos\left(\left|\zeta\right|+\zeta_{0}\right)\,\sinh\left[ q\,\left(\left|\zeta\right|+\zeta_{0}-\pi\right)\right]-q\sin\left(\left|\zeta\right|+\zeta_{0}\right)\,\cosh\left[q\,(\left|\zeta\right|+\zeta_{0}-\pi) \right] \right\}.\nonumber
\end{eqnarray}
In particular, the function $f$ on the brane takes the form
\begin{equation}
\label{final shock wave for ads}
f(\zeta=0,\chi)=-\frac{p}{2\pi M_{4}^2}\,\int_{0}^{\infty} dq \frac{\cos(q\chi)}{1+q^2}\left[\cot \zeta_{0}+q\coth\left[q(\pi-\zeta_0) \right] \right] \mbox{ .}
\end{equation}
Using the theorem of residues, we can show that for $\zeta_0 \ne 0$, eq.~(\ref{final shock wave for ads}) can also be written as
\begin{equation}\label{fseries}
f\left(\zeta=0,\,\chi\right)=-\frac{p\,\alpha^2}{2\pi M_{4}^2}\,\sum_{n=1}^{\infty}\frac{n}{n^2\,\alpha^2-1}\,e^{-n\,\alpha\left|\chi\right|} \mbox{ ,} 
\end{equation}
where we have defined the dimensionless quantity
\begin{equation}
\alpha\equiv\frac{\pi}{\pi-\zeta_0}\,.
\end{equation}

We can now check the validity of this result by considering some special limits where the shock wave metric is already known.

\subsection{Two limits: flat brane and no bulk}

The limit of a flat brane ($\ell_{3}\rightarrow \infty$) is obtained by sending $\zeta_0\rightarrow 0$. In this case we find from (\ref{final shock wave for ads})
\begin{eqnarray}
\label{l3 goes to infinity limit}
f(\zeta=0,\,\chi)=-\frac{p}{4\pi M_{4}^2}\left[\pi\,e^{-\left|\chi\right|}\cot \zeta_0|_{\zeta_0\rightarrow 0}- \left|\chi\right|\,e^{-\left|\chi\right|}-1\right. &- & \left. 2\,\cosh\left|\chi\right|\log(1-e^{-\left|\chi\right|}) \right]+\nonumber\\
&&+{\cal {O}}\left(\zeta_0\right),
\end{eqnarray}
where $\cot \zeta_{0}=\sqrt{\ell_{3}^{\,2}-\ell_{4}^{\,2}}/\ell_{4}$. 

In order to deal with the divergent term $\cot \zeta_0|_{\zeta_0\rightarrow 0}$ we set $\chi=R/\ell_{3}$ (where $R$ is the proper radial distance from the source) before sending $\ell_3\rightarrow\infty$ while keeping $R$ finite. In this limit the metric reads
\begin{equation}
\label{flat brane limit}
f(\zeta=0,\,\chi)=\frac{p}{4\pi \ell_{4} M_{4}^2}\left[\pi \left|R\right|+\ell_{4}\log(R/\ell_{3})^2 \right] \mbox{ ,}
\end{equation}
plus a divergent term ($\propto \ell_3$) that is however independent of the coordinates and can be set to zero by a coordinate transformation.

The first term in~(\ref{flat brane limit}) is the contribution from $2+1$ gravity and is associated to the deficit angle generated by a mass in $2+1$ dimensions~\cite{Ferrari:1988cc,Barrabes:2002hn}, while the second term has the same form as a $3+1$ dimensional shock wave~\cite{Aichelburg:1970dh}. Eq.~(\ref{flat brane limit}) coincides with the result previously found in~\cite{Emparan:2001ce}.  Comparing the above result~(\ref{flat brane limit}) with the known $2+1$ dimensional Minkowski shockwave ($f= p\left|R\right|/2M_3^2$~\cite{Ferrari:1988cc}), we obtain the expression of the effective $2+1$ dimensional Planck mass {\em for a flat brane}: $M_3^{\ell_3\rightarrow \infty}=2\,\ell_4\,M_4^2$.

\smallskip

One more check can be made by taking the limit in which the bulk disappears. This is achieved by sending $\ell_4 \rightarrow 0$ (i.e. $\zeta_0\rightarrow 0$) and $M_4^{\,2} \rightarrow \infty$ while keeping the product $2\,\ell_4\,M_4^{\,2}=M_3^{\ell_4\rightarrow 0}$ finite. In this limit we find $\cot \zeta_0 \simeq \ell_3/\ell_4$, and using (\ref{l3 goes to infinity limit}) we obtain
\begin{equation}
f(\zeta=0,\,\chi)=-\frac{\ell_3\,p}{2\,M_3^{\ell_4\rightarrow 0}}e^{-\left| \chi\right|}\mbox{,}
\end{equation}
which, as expected, matches~(\ref{shockads3}) once we set $M_3^{\ell_4\rightarrow 0}=M_3$.

\subsection{Mass spectrum}

Following a procedure analogous to that of~\cite{Kaloper:2005wq}, we can find the mass spectrum for the graviton. In order to do this, we decompose the shock wave in Kaluza-Klein modes by writing (\ref{psiequationads}) as
\begin{equation}
\label{psi equation ads}
\partial_{\zeta}^2\psi_{q}-2\cot\left(\left| \zeta\right| +\zeta_0 \right)\partial_{\left|\zeta\right|}\psi_{q}=-m^2\ell_3^2\,\psi_q \mbox{ ,}
\end{equation}
where $m^2\ell_3^2=-1-q^2$. Here, the values of $q$ are given by the poles in (\ref{final shock wave for ads})
\begin{equation}
 q_{n}=i\frac{\pi\,n}{(\pi-\zeta_0)}\:\:,\qquad n=1,\,2,\,3,\,\ldots \mbox{ .}
\end{equation}
Thus our result is
\begin{equation}\label{kkmasses}
m_{n}^2\,\ell_3^2=-1+\frac{\pi^2\,n^2}{(\pi-\zeta_{0})^2}\mbox{ .}
\end{equation}

We see that, analogously to the Karch-Randall case~\cite{Karch:2000ct} of a $AdS_4$ brane in $AdS_5$ bulk, there is no zero mode of the graviton as long as $\zeta_0\neq 0$. Moreover, this technique allows us to prove that the mechanism that leads to an ultralight graviton is at work also in the lower dimensional case. In the case of a $AdS_4$ brane in a $AdS_5$ bulk, indeed, the lightest mode of the graviton goes as $\sqrt{3/2}\, \ell_5/\ell_4^2\,\left(1+{\cal {O}}\left(\ell_5/\ell_4\right)\right)$ for $\ell_5\ll \ell_4$ ~\cite{Karch:2000ct}. In our case of a $AdS_3$-brane, for $\zeta_0\ll 1$, the mass of the ultralight mode reads $m_{\mathrm {UL}}^2\simeq 2\,\zeta_0/\pi\ell_3^2$, i.e. 
\begin{equation}\label{massul}
m_{\mathrm {UL}}\simeq\sqrt{\frac{2\,\ell_4}{\pi\,\ell_3}}\,\frac{1}{\ell_3}\left(1+{\cal {O}}\left(\frac{\ell_4}{\ell_3}\right)\right)\,\,.
\end{equation}
The scaling $m_{\mathrm {UL}}\propto \ell_4^{1/2}/\ell_3^{3/2}$ is in agreement with expectations from AdS/CFT arguments \cite{Porrati:2001gx,Porrati:2001db,Porrati:2003sa,Duff:2004wh}. Indeed, since the phenomenon of mass generation for the graviton is associated to gravitational dynamics, we must have $m_{\mathrm {UL}}^2\propto 1/M_3$. Since $\ell_3$ is the only other scale on the CFT side of the problem, then, $m_{\mathrm {UL}}^2\propto 1/M_3\,\ell_3^3$. Finally, $m_{\mathrm {UL}}^2\propto g_*$ where $g_*\propto M_4^2\,\ell_4^2$ is the number of degrees of freedom in the CFT. Putting  together these factors (and using $M_3\propto M_4^2\ell_4$) we readily recover the scaling~(\ref{massul}).

Following an argument analogous to that of~\cite{Kaloper:2005wq}, we can also find the value of the effective $2+1$ dimensional Planck mass for finite $\ell_3$, defined as the coupling to matter of the ultralight mode of the graviton. Comparing~(\ref{fseries}) to~(\ref{shockads3}) we find 
\begin{equation}\label{masseff}
M_3\left(\ell_3\right)=2\,M_4^2\ell_3\,\zeta_0\,\left(1-\frac{\zeta_0}{2\pi}\right)\simeq 2\,M_4^2\ell_4\,\left[1-\frac{1}{2\pi}\left(\frac{\ell_4}{\ell_3}\right)+{\cal {O}}\left(\left(\frac{\ell_4}{\ell_3}\right)^2\right)\right]\,.
\end{equation}

\subsection{The CFT energy momentum tensor}

One of the main motivations for the study of the gravitational field of localized sources in the Randall-Sundrum model is the dual interpretation of such solutions as quantum-corrected metrics~\cite{Tanaka:2002rb,Emparan:2002px}. Once the brane metric is given, it is straightforward to derive the $2+1$ dimensional stress-energy tensor that supports it. Such tensor contains {\em two} contributions: the first is just the stress energy tensor of the original, classical sources. A second contribution is associated to gravitational backreaction of the CFT modes that are excited by the gravitational field of the classical source.

The general expression for the $2+1$ energy momentum tensor is derived in the appendix, eq.~(\ref{induced energy momentum tensor}). For the shock wave~(\ref{final shock wave for ads}) we find that the expression for the CFT tensor reads
\begin{equation}\label{cftdray}
T_{uu}^{\mathrm {CFT}}=-\frac{2\,M_{3}}{\ell_3^{\,2}}\,\delta(u)\,\left[1-\frac{\partial^{\,2}}{\partial \chi^{2}} \right]\,f(\chi) =-\frac{\alpha^2}{4\pi}\,\frac{M_{3}}{M_{4}^{2}\,\ell_3^{\,2}}\,p\,\frac{1}{\sinh^2(\alpha\chi/2)}\,\delta(u)
\end{equation}
where we see that, as a consequence of Lorentz contraction, the stress energy tensor of the CFT is also localized on the null surface $u=0$. At variance with the stress energy tensor of the classical source, however, the CFT tensor has a nontrivial profile along the $\chi$ coordinate. 

The effect of the CFT in the solution~(\ref{final shock wave for ads}) can also be interpreted as a running of the effective Planck constant~\cite{Kaloper:2005wq}. In this case, the expression~(\ref{masseff}) corresponds to the infrared limit of the $2+1$ dimensional Planck mass, and its running is induced by the effect of the Kaluza-Klein modes in eq.~(\ref{fseries}).

\section{Boosting the AdS C-metric}

Let us now turn to the second way of getting a shock wave metric, i.e. by boosting to the speed of light the gravitational field of a massive particle while sending to zero the mass of the particle, so that the particle momentum stays finite. 

\subsection{The AdS C-metric}
\FIGURE[left]{\epsfig{file=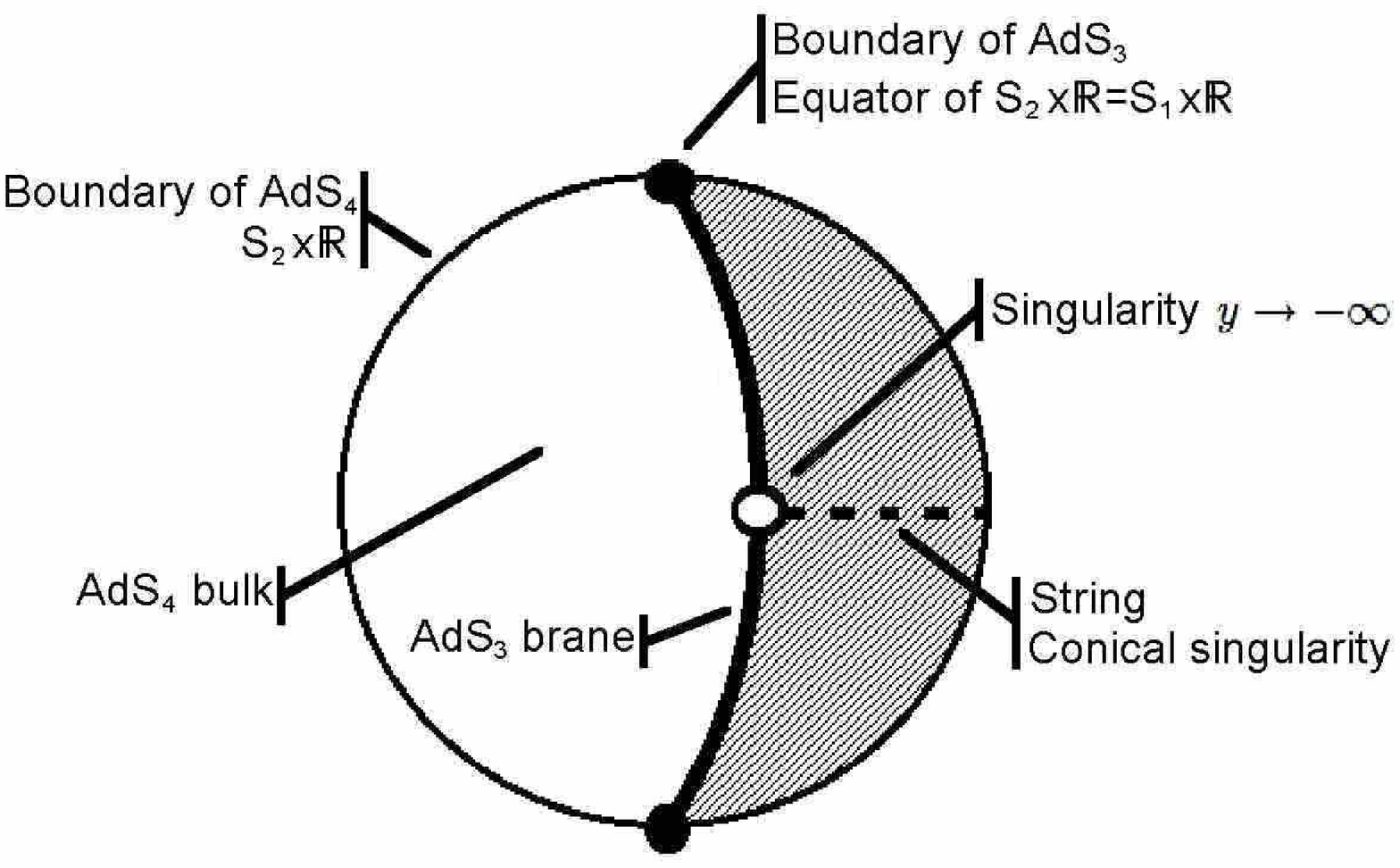, width=8cm, height=5cm}
\caption{The Karch-Randall black hole construction of~\cite{Emparan:1999fd}: we show here a constant time slice of $AdS_4$; the singularity at $y\rightarrow  -\infty$ [see eq.~(\ref{adsc})] corresponds to the location of the particle, the dashed line describes the string that is pulling the particle towards the $AdS_4$ boundary. The shaded area corresponds to the region that is suppressed by the cut-and-paste Randall Sundrum procedure.} \label{figure}}

Our starting point is the $AdS_3$-brane black hole solution given in~\cite{Emparan:1999fd}. In that paper, a black hole solution is found by observing that a particle on the Randall-Sundrum (or Karch-Randall) brane has to be accelerated with respect to the bulk. In order to generate such an acceleration, it is necessary to attach the particle to a string pulling it towards the $AdS$ boundary (we sketch this construction in figure 1). The metric corresponding to this setup is known in four dimensional $AdS$ space\footnote{For a study of analogous metrics in higher dimensions see~\cite{Charmousis:2003wm}.} and is called {\em AdS C-metric}~\cite{Plebanski:1976gy}. In order to construct a brane black hole solution it is then sufficient to cut the bulk $AdS$ space with a brane at the location of the particle, throwing away the part of space that contains the string, and gluing to the brane a $Z_2$-symmetric copy of the part of $AdS$ space that has been retained.

To obtain the brane shock wave associated to the black hole solution of~\cite{Emparan:1999fd}, we will eventually boost the whole system of string+particle in a direction that is transverse to that of the string, in such a way that the particle remains on the surface of the brane, while sending the mass of the particle to zero.

Our starting point is the $AdS$ C-metric, that can be  written in the form
\begin{eqnarray}\label{adsc}
ds^2&=&\frac{1}{A^2\,\left(x-y\right)^2}\left[-H\left(y\right)\,dt^2+\frac{dy^2}{H\left(y\right)}+\frac{dx^2}{G\left(x\right)}+G\left(x\right)\,d\phi^2\right]\,\,,\nonumber\\
H\left(y\right)&=&\lambda-k\,y^2+\frac{mA}{4\pi\,M_4^2}\,y^3\,\,,\qquad
G\left(x\right)=1+k\,x^2-\frac{mA}{4\pi\,M_4^2}\,x^3\,\,,
\end{eqnarray}
with $\lambda>0$ and $k=-1,\,0,\,+1$. Here $m$ is interpreted (at least in the small $m$ regime) as the mass of the particle and $A$ is its acceleration. 

The brane is located at $x=0$~\cite{Emparan:1999fd}, so that the brane induced metric reads (after some simple coordinate redefinition, and setting $\lambda\,A^2\equiv 1/\ell_3^2$)
\begin{equation}\label{adscproj}
ds^2=-\left(\frac{r^2}{\ell_3^2}-k+\frac{m}{4\pi\,M_4^2\, r}\right)\,dt^2+\left(\frac{r^2}{\ell_3^2}-k+\frac{m}{4\pi\,M_4^2\, r}\right)^{-1}dr^2+r^2\,d\phi^2
\end{equation}
where the periodicity of $\phi$ depends on $m$ as described below. Our first task is to choose the sign of $k$. To do this, we consider the fact we will eventually send the parameter $m$ to zero while boosting the source to the speed of light. In order for the boost to be well defined, we need the background coordinate system to cover  the whole $AdS_3$. This forces us to choose the branch with $k=-1$, since both branches $k=0$ and $k=+1$ have a horizon at $r=\ell_3\,\sqrt{k}$ and do not cover all of $AdS_3$.

So from now on we will set $k=-1$ in the metric~(\ref{adsc}). For a detailed interpretation of this metric see e.g.~\cite{Podolsky:2002nk,Dias:2002mi,Krtous:2005ej}. For the present work, all we need to know is that $-1/y$ is a radial coordinate from the particle (that is located at a singularity at $y\rightarrow -\infty$) while $x$ is roughly interpreted as $\cos\theta$ in polar coordinates. The $x$ coordinate is bound to be larger than $y$, and the $AdS_4$ boundary is the surface $x=y$.  The equation $G\left(x\right)=0$ has three roots, out of which only one (for $m>0$) is positive, let us call it $x_2$. The fact that $x$ is interpreted as $\cos\theta$ implies that $x=x_2$ corresponds to a polar axis. Since we will eventually introduce the Randall-Sundrum brane at $x=0$, cutting away the region $x<0$, we only care about the region $0<x<x_2$. In order to avoid a conical singularity on the axis $x=x_2$ then we impose that the angle $\phi$ ranges between $0$ and $4\pi/\left|G'\left(x_2\right)\right|\simeq 2\pi\left[1-mA/4\pi M_4^2+{\cal O}\left((mA/M_4^2)^2\right)\right]$.

For $m=0$ the above metric~(\ref{adsc}) describes empty $AdS_4$. To see this, we remind that $AdS_4$ can be embedded in five dimensional space with coordinates $W_0,\,W_1,\,W_2,\,W_3,\,W_4$, as discussed in section 3. In terms of these coordinates, the metric (\ref{adsc}) with $m=0$ is given by
\begin{eqnarray}
&&W_0=\frac{1}{A\sqrt{\lambda}}\,\frac{\sqrt{y^2+\lambda}}{x-y}\,\cos\left(\sqrt{\lambda}\,t\right)\,\,,\nonumber\\
&&W_1=\frac{1}{A\sqrt{\lambda+1}}\,\frac{y+\lambda\,x}{\sqrt{\lambda}\,\left(y-x\right)}\,\,,\nonumber\\
&&W_2=\frac{1}{A}\,\frac{\sqrt{1-x^2}}{x-y}\,\cos\phi\,\,,\nonumber\\
&&W_3=\frac{1}{A}\,\frac{\sqrt{1-x^2}}{x-y}\,\sin\phi\,\,,\nonumber\\
\label{embedding ads4}
&&W_4=\frac{1}{A\sqrt{\lambda}}\,\frac{\sqrt{y^2+\lambda}}{x-y}\,\sin\left(\sqrt{\lambda}\,t\right)\,\,.
\end{eqnarray}

The $AdS_4$ radius is
\begin{equation}
\ell_4=\frac{1}{A\,\sqrt{\lambda+1}}\,\,.
\end{equation}

For finite $m$, the metric has a singularity at $y\rightarrow -\infty$, that corresponds, in the limit of small $m$, to a particle whose worldline follows
\begin{eqnarray}
&&\bar{W}_0=\ell_4\frac{\sqrt{\lambda+1}}{\sqrt{\lambda}}\,\cos\left(\sqrt{\lambda}\,t\right)\,\,,\nonumber\\
&&\bar{W}_1=\frac{\ell_4}{\sqrt{\lambda}}\,\,,\nonumber\\
&&\bar{W}_2=\bar{W}_3=0\,\,,\nonumber\\
\label{coordinates with m=0}
&&\bar{W}_4=\ell_4\frac{\sqrt{\lambda+1}}{\sqrt{\lambda}}\,\sin\left(\sqrt{\lambda}\,t\right)\,\,.
\end{eqnarray}

The $AdS_3$ brane will eventually be located at $x=0$~\cite{Emparan:1999fd}, corresponding to the plane $W^{\mathrm {brane}}_1=\ell_4/\sqrt{\lambda}$. By comparing this with expression~(\ref{ads3inads4}), or equivalently by looking at~(\ref{adscproj}),  we find that the $AdS_3$ curvature of the brane will be given by $1/\ell_3=A\,\sqrt{\lambda}$.

\subsection{The boosted AdS C-metric}

We now perform a boost along the $\left(W_0,\,W_2\right)$ direction (so that the $W_1$ coordinate, and therefore the location of the brane, is left unchanged), by replacing
\begin{eqnarray}
&&W_0\rightarrow \gamma\,\left(W_0-\beta\,W_2\right)\,\,,\nonumber\\
&&W_2\rightarrow \gamma\,\left(W_2-\beta\,W_0\right)\,\,,
\end{eqnarray}
with $\gamma=1/\sqrt{1-\beta^2}$. Finally, we take the limit $\beta\rightarrow 1,\,m\rightarrow 0$ with $m\,\gamma=p$ finite.

To find the boosted result we use the same procedure described in e.g.~\cite{Hotta:1992qy}: first, we expand the metric at first order in $m$ and we replace $m\rightarrow p/\gamma$, then we send $\gamma\rightarrow \infty$ while using the identity
\begin{equation}
\lim_{\beta\rightarrow 1}\,\gamma\,f\left(\gamma\left(W_0-\beta\,W_2\right)\right)=\delta(W_0-W_2)\int_{-\infty}^{\infty}f\left(w\right)dw\mbox{ ,}
\end{equation}
that can be easily proved by treating $\gamma\,f\left(\gamma\left(W_0-\beta\,W_2\right)\right)$ as a distribution~\cite{Hotta:1992qy}.

In order to boost properly our metric, we have to take into account also the deficit angle in $\phi$. In the limit $m\rightarrow 0$, we have indeed $0<\phi\la 2\pi\left(1-mA/4\pi M_4^2\right)$, so that before expanding at first order in $m$ we have to perform the redefinition $\phi\rightarrow \phi/\left(1-mA/4\pi M_4^2\right)$. This way, $\phi$ ranges on its natural interval $[0,\,2\pi[$.

The final result is
\begin{eqnarray}\label{boostedemb}
ds^2&=&ds_{\mathrm {AdS_4}}^2+\frac{p\,A}{\pi\,M_{4}^2}\,\ell_4\,\left(\lambda+1\right)\,\delta\left(W_0-W_2\right)\,d\left(W_0-W_2\right)^2\\
&&\times\left[-\sqrt{\lambda+1}+\frac{W_4}{\ell_4}\,\sqrt{\lambda}\,{\mathrm {tanh}}^{-1}\left(\frac{\sqrt{\lambda}\,\ell_4+W_1}{\sqrt{\lambda+1}\,W_4}\right)-\frac{W_3}{\ell_4}\,{\mathrm {tan}}^{-1}\left(\frac{\sqrt{\lambda+1}\,W_3}{\ell_4-\sqrt{\lambda}\,W_1}\right)\right]\,.\nonumber
\end{eqnarray}

This is one of the main results of our paper: it represents the metric (in the embedding coordinates $W_0,\,...,\,W_4$) describing a null particle in $AdS_4$ with radius $\ell_4$ subject to an acceleration $A=1/\ell_4\sqrt{\lambda+1}$ transverse to the direction of motion of the particle.

We now obtain the shock wave metric on the $AdS_3$ brane by using the coordinate system~(\ref{ads3inads4}), where we set $z=z_0=\ell_3\,\zeta_0$ so that $\sin\zeta_0=\ell_4/\ell_3$, and then using for the $Z_0,\,...,\,Z_3$ coordinates the light cone system~(\ref{ads3lightcone}). The resulting metric is
\begin{eqnarray}\label{finalboost}
ds^2=ds_{\mathrm {Empty\ AdS_3}}^2&&-\frac{p}{\pi\,M_4^{2}\cos^2 \zeta_0}\times\\
&&\times\left[\pi\,\cot \zeta_0\,\sinh\left|\chi\right|+\left(2-\cosh\chi\,\log\frac{\cosh\chi+1}{\cosh{\chi}-1} \right) \right]\delta(u)du^2.\nonumber
\end{eqnarray}
We can bring this result to a form that makes comparison with the pure $AdS_3$ case easier. Following~\cite{Cai:1999dz}, we write the metric~(\ref{finalboost}) in terms of the embedding coordinates~(\ref{ads3lightcone}), so that it reads
\begin{eqnarray}\label{boostemb}
ds^2=&&-dU\,dV+dZ_2^2-dZ_3^2+\\
&&-\frac{p}{2\pi\,M_4^{2}\cos^2 \zeta_0}\,\left[\pi\,\cot \zeta_0\,\frac{\left|Z_2\right|}{\ell_3}+\left(2-\frac{Z_3}{\ell_3}\,\log\frac{Z_3+\ell_3}{Z_3-\ell_3} \right) \right]\delta(U)\,dU^2 \mbox{ ,} \nonumber
\end{eqnarray}
with $U=Z_0-Z_1$ and $V=Z_0+Z_1$. We then perform the transformation
\begin{eqnarray}
&&U\rightarrow U\,\,,\nonumber\\
&&V\rightarrow V-\frac{p}{2\,M_4^2}\,\frac{\cot \zeta_0}{\cos^2 \zeta_0}\,\frac{Z_3}{\ell_3}\,\Theta\left(U\right)-\left(\frac{p}{4\,M_4^2}\,\frac{\cot \zeta_0}{\cos^2 \zeta_0}\right)^2\,\frac{U}{\ell_3^2}\,\Theta\left(U\right) \,\,,\nonumber\\
&&Z_2\rightarrow Z_2\,\,,\nonumber\\
&&Z_3\rightarrow Z_3+\frac{p}{4\,M_4^2}\,\frac{\cot \zeta_0}{\cos^2 \zeta_0}\,\frac{U}{\ell_3}\,\Theta\left(U\right)\,\,,
\end{eqnarray}
where $\Theta$ is the Heaviside step function. This transformation has the effect of replacing $\left|Z_2\right|$ with $\left|Z_2\right|-Z_3$ in equation~(\ref{boostemb}), or equivalently to bring~(\ref{finalboost}) to the form
\begin{equation}\label{finalboost1}
ds^2=ds_{AdS_3}^2+\frac{p}{\pi\,M_4^{2}\cos^2 \zeta_0}\,\left[\pi\,\cot \zeta_0\,e^{-\left|\chi\right|}-\left(2-\cosh\chi\,\log\frac{\cosh\chi+1}{\cosh{\chi}-1} \right) \right]\delta(u)du^2 \mbox{ .}
\end{equation}

This is the final expression for the brane metric obtained by boosting the solution~(\ref{adsc}). The first term in the above metric is the contribution from the deficit angle. It corresponds to the term in $e^{-\left|\chi\right|}$ in eq.~(\ref{shockads3}). The second term has the same form of the $AdS$ shock wave in $3+1$ dimensions~\cite{Podolsky:1997ri}. Its presence reflects the fact that the black hole solution of~\cite{Emparan:1999fd} contains one term that corresponds to the classical $AdS_3$ conical singularity and a "quantum" term that resembles that of the $3+1$ dimensional $AdS$ black hole and dresses the conical singularity with a horizon.

As a check of the validity of the result above, it is straightforward to show that the limits of a flat brane and of no bulk give the same resulting brane metric as that shown in section 3.1.

\subsection{The CFT energy momentum tensor}

By using the procedure described in subsection 3.3, we can find the energy momentum tensor associated to the CFT
\begin{equation}\label{cftboost}
T_{uu}^{\mathrm {CFT}}=-\frac{1}{\pi\,\cos^2 \zeta_0}\,\frac{M_3}{M_4^2\,\ell_4^2}\,p\,\frac{1}{\sinh^2\chi}\,\delta\left(u\right)\,\,.
\end{equation}
This expression can be found also by boosting the energy momentum tensor~\cite{Emparan:2002px} associated to the CFT around the brane-localized black hole of ~\cite{Emparan:1999fd}
\begin{equation}
T^{\alpha}_{\,\,\beta}\propto\frac{1}{r^3}{\mathrm {diag}}\left(1,\, 1,\, -2\right)\mbox{ .}
\end{equation}

The expression~(\ref{cftboost}) is different from the induced CFT stress energy tensor found in section 3.3, but its form is extremely similar to it, the only difference being on the dependence on $\chi$ rather than on $\alpha\,\chi/2$ (and indeed the two expressions coincide for $\zeta_0=\pi/2$, i.e. $\alpha=2$).

It is worth remarking that also for this solution one can interpret the effect of the CFT as running of the effective Planck mass. In this case, the running of the Planck scale appears to be different from that observed in section 3.

\section{Discussion: two shock waves on the $AdS_3$ brane}

We have constructed the metric associated to a null source on a $AdS_3$ brane embedded in a $AdS_4$ bulk in two different ways. 

In section 3 we have obtained this metric by directly solving Einstein equations for a null source.  The resulting brane shockwave is given in equation~(\ref{fseries}). In order to obtain this result, we have effectively decomposed the shock wave in Kaluza-Klein modes, imposing that the solution is regular at the $AdS_4$ boundary of our bulk. This allowed us to find the expression for the masses of the Kaluza-Klein  graviton~(\ref{kkmasses}) and to prove that, similarly to the case of a $AdS_4$ brane in $AdS_5$ bulk, there is no zero mode of the graviton, even if there is an ultralight mode, whose Compton wavelength is much larger than $\ell_3$. The reason for the absence of a zero mode was discussed in~\cite{Karch:2000ct}: an observer on the $AdS_3$ brane can see all of the $AdS_4$ bulk, including its boundary. The "would be" zero mode of the graviton is divergent (and non normalizable) on the bulk $AdS_4$ boundary, and therefore decouples from the brane matter. The absence of a massless mode of the graviton is confirmed by the fact that for large $\chi$ the shockwave goes as $e^{-\alpha\left|\chi\right|}$, with $\alpha=\pi/\left(\pi-\zeta_0\right)>1$ whereas for a theory of massless gravity in $AdS_3$ one expects $f\propto e^{-\left|\chi\right|}$.

In section 4, then, we have obtained a shock wave by boosting to the speed of light the $AdS_3$ brane black hole of~\cite{Emparan:1999fd} while sending its mass to zero. The resulting brane metric is given in~(\ref{finalboost1}). It is apparent that the expressions~(\ref{fseries}) and~(\ref{finalboost1}) do not coincide! In particular, the boost of the brane black hole metric of~\cite{Emparan:1999fd} seems to excite a massless graviton, since at large distances $f\propto e^{-\left|\chi\right|}$.

Where does this different behavior come from? The answer lies in the different behavior on the $AdS_4$ boundary. In the case discussed in section 3, we have imposed by hand that the function $f$ goes to zero as $\zeta\rightarrow \pi-\zeta_0$. A section of the metric given by~(\ref{shockkr}) and~(\ref{fkr}) at $\zeta=\pi-\zeta_0-\epsilon$ (that in the absence of shock wave gives an $AdS_3$ space) gives an induced metric that is that of $AdS_3$ with radius $\ell\simeq \ell_4/\epsilon$ with a correction associated to the brane shockwave that is proportional to $\epsilon^3$:
\begin{equation}\label{draybound}
ds^2_{\mathrm {eqs~(\ref{shockkr},\ref{fkr})}}\left.\right|_{\zeta=\pi-\zeta_0-\epsilon}=\frac{1}{\epsilon^2}\,\left[ds^2_{\mathrm {Empty AdS_3}}+{\cal{O}}\left(\epsilon^3\right)\right]\,\,.
\end{equation}

On the other hand, the boosted metric~(\ref{boostedemb}), cut at a surface $z=\left(\pi-\epsilon\right)\,\ell_3$ (equivalent to $\zeta=\pi-\zeta_0-\epsilon$), reads, for $\epsilon\ll 1$,
\begin{eqnarray}\label{boostbound}
&&ds^2_{\mathrm {eq~(\ref{boostedemb})}}\left.\right|_{\zeta=\pi-\zeta_0-\epsilon}=\frac{1}{\epsilon^2}\,\left\{ds^2_{\mathrm {Empty AdS_3}}+\frac{2\,p\,A\,\ell_4^2}{\pi\,M_4^2\,\ell_3}\,\left(\lambda+1\right)\,\times\right.\nonumber\\
&&\left.\times\left[\sqrt{\lambda}\,\cosh\chi\,{\mathrm {tanh}}^{-1}\left(\frac{1}{\sqrt{\lambda+1}\cosh\chi}\right)+\sinh\chi\,{\mathrm {tan}}^{-1}\left(\frac{\sqrt{\lambda+1}\sinh\chi}{\sqrt{\lambda}}\right)\right]\delta\left(u\right)du^2+{\cal{O}}\left(\epsilon\right)\right\},\nonumber\\
\end{eqnarray}
that contains a nontrivial term as $\epsilon\rightarrow 0$. It is possible to verify that eq.~(\ref{boostedemb}) is actually a solution to the shock wave equation~(\ref{eqf}). However, such a solution cannot be decomposed \`a la Kaluza-Klein, since the individual modes that make this solution are not square--summable close to the $AdS_4$ boundary.

We conclude that both eqs.~(\ref{shockkr},\ref{fkr}) and eq.~(\ref{boostedemb}) give a legitimate shock wave on the brane, however they have radically different behavior at the bulk boundary. We now turn to the CFT interpretation of this phenomenon.

\subsection{The CFT interpretation}

It is of course interesting to interpret the existence of these two different shock wave solutions in terms of the CFT dual of the Karch-Randall model. 

Subsequently to the paper~\cite{Karch:2000ct}, that first observed the absence of a zero mode for the graviton in this setting, it has been shown~\cite{Porrati:2001gx, Porrati:2001db,Porrati:2003sa,Duff:2004wh} that the ultralight mass for the graviton can be  generated by a CFT with appropriate ({\em {transparent}}) boundary conditions on a $AdS$ background. The shock wave of section 3 displays the properties that we expect to find in massive gravity.

On the other hand, the existence of a $AdS$ black hole solution of~\cite{Emparan:2002px} associated to the $AdS$ C-metric should also be explained by the existence of a CFT with transparent boundary conditions on the boundary of $AdS_3$. This has not been shown in a rigorous way, since there is no explicit computation of the stress energy tensor of a CFT around a conical singularity in $AdS_3$. However, the $k=-1$ branch of eq.~(\ref{adscproj}) (that we have used to generate the shock wave of section 4) can be continuously deformed into the branch with $k=+1$, that can be shown to correspond to a BTZ black hole~\cite{Banados:1992wn} corrected by a CFT with transparent boundary conditions~\cite{Steif:1993zv,Lifschytz:1993eb,Shiraishi:1993hf}.

So it seems that both the shock waves of section 3 and 4 are obtained by endowing the {\em same} source (a null particle in $AdS_3$) with the quantum corrections of a CFT with the {\em same} (transparent) boundary conditions. Now: why the two solutions are different? Here we argue that the difference between these solutions emerges from the fact that the dual of the Karch-Randall model contains actually {\em two} CFTs. The two different solutions corresponds to situations where the second CFT is in a different state.

The CFT structure of the Karch-Randall model is more complicated than that of a simple CFT in interaction with gravity. Indeed, the CFT ``on the brane'' does not describe all of the bulk degrees of freedom, but only those degrees of freedom that lie in the holographic domain of the brane, i.e. whose holographic projection lies on the brane. As discussed in~\cite{Bousso:2001cf}, for a brane located at $z=0$ in the coordinates~(\ref{ads4slicedads3new}), the holographic domain corresponds to the region $0<z<\pi \ell_3/2$. The remaining part of bulk is mapped into a CFT that lives on the boundary of $AdS_4$. The two CFTs communicate though the common boundary of the spaces where they live (the equator of $S_2\times \Re$ of figure 1).

We can now associate the shock wave described in section 3 to the case where the second CFT is in its ground state. Indeed, the metric~(\ref{draybound}) induced on the fictitious brane at $z=\left(\pi-\epsilon\right)\,\ell_3$ is pure $AdS_3$ metric for $\epsilon\rightarrow 0$. This corresponds to putting the second CFT in its vacuum state. We stress that this is the situation where the dynamical generation of a mass for the graviton is observed. 

The metric~(\ref{boostbound}), on the other hand, gives a nontrivial metric at the $AdS_4$ boundary, that corresponds to a deformation of the second CFT. It is interesting to notice that in this case the long distance behavior of the shockwave corresponds to the one associated to a massless graviton. A possible interpretation is that the deformation of the boundary conditions leaves the second CFT in an excited state, effectively changing the boundary conditions of our brane CFT. Since the boundary conditions of the CFT living in the $AdS_3$ space are crucial in determining whether the graviton mass is generated (see e.g.~\cite{Porrati:2001db}), it is natural to imagine that this different state prevents the gravitational Higgs phenomenon from taking place. 

At this point, it is also important to note that the original construction of~\cite{Emparan:1999fd} contained {\em two branes}. The second brane was introduced to insure that the graviton spectrum contains a zero mode (however, in the spirit of the Karch-Randall model, the presence of this second brane is not necessary).  Remarkably, the second brane was located at $z=\pi \ell_3/2$ in the coordinates~(\ref{ads4slicedads3new}), that is exactly at the boundary of the holographic region found in~\cite{Bousso:2001cf}. This allows us to find a second CFT interpretation of the brane black hole of~\cite{Emparan:1999fd} and of the shock wave found in section 4. In this interpretation, the second CFT does not exist at all (since the corresponding part of bulk has been thrown away), and the graviton remains massless.

These results show explicitly how different metrics can be obtained when we deform the second CFT. More in general, they show that the choice of boundary conditions in the CFT can affect strongly the nature of the quantum corrected metrics of localized objects. They also raise a natural question: what happens if we "unboost" the metric found in section 3? In other words, suppose that now we see the shock wave geometry as the metric of an extremely light but not massless particle. In this case one can go to the rest frame for this particle, and ask what the metric will look like. Since we have found two shock waves, it is natural to expect the existence of two branches of solutions associated to finite mass, brane localized objects. It would be very interesting to study the nature of the objects belonging to the second branch.

\vskip0.5cm

To sum up, in this paper we have seen the power of gravitational shock waves in the study of the properties of brane gravity, by showing that there are (at least) {\em two} different solutions to Einstein equations for a null source moving along a $AdS_3$ brane embedded in a $AdS_4$ bulk. From the bulk perspective, the origin of these two different solutions is clearly explained by different conditions at the $AdS_4$ boundary of the bulk. In particular, one of the shock waves excites a profile of the bulk graviton that is not normalizable from the three-dimensional point of view. The CFT interpretation of these two solution is more subtle, and we argued that the two different solutions correspond to putting the $AdS_4$ boundary CFT in a different state. 

\smallskip

{\bf \noindent Acknowledgements}

\smallskip

It is a pleasure to thank Sergei Dubovsky, David Kastor, Marco Peloso, Jennie Traschen and especially Nemanja Kaloper for useful discussions. 

\appendix
\section{ Shock wave on $AdS_{n-1}$ brane embedded in $AdS_{n}$ space}

In this appendix we work out all the relevant formulae needed for the study of gravitational shock waves associated to particles localized on a $AdS_{n-1}$ brane embedded in $AdS_{n}$ space.

Consider the following metric $d\tilde s^{\,2}_n=\Omega^2(z)ds_n^2$ where
\begin{equation}
\label{ general metric in n dimension}
ds_n^2=2\,A(u,v)\,dudv+g\left(u,\,v\right)\,h_{ij}(x)\,dx^idx^j+dz^2\mbox{ ,}
\end{equation}
where $i,\,j=1,\,2,\,...,\,n-3$, and $u$ and $v$ are null coordinates. We also assume that there exist matter fields with energy momentum tensor given by
\begin{eqnarray}
\nonumber
\tilde T&=&2\,\tilde T_{uv}(u,v,x,z)\,dudv+ \tilde T_{uu}(u,v,x,z)\,du^2+ \tilde T_{vv}(u,v,x,z)\,dv^2+ \\
&&+\tilde T_{ij}(u,v,x,z)\,dx^idx^j\label{energy momentum in general}
+\tilde T_{zz}(u,v,x,z)\,dz^2 \,\,.
\end{eqnarray}
 Now consider a massless particle located at $u=0$ and moving with the speed of light in the direction of $v$.  Dray and 't Hooft showed that the effect of this particle on the background geometry can be described by the metric (\ref{ general metric in n dimension}) and the energy momentum tensor (\ref{energy momentum in general}) for $u<0$ and by making the shift $v \rightarrow v+f(x,z)$ in (\ref{ general metric in n dimension}) and (\ref{energy momentum in general}) for $u>0$. The resulting metric and the energy momentum tensor read \cite{Dray:1984ha,Sfetsos:1994xa}
\begin{eqnarray}
\nonumber
ds_n^2&=&2\,A(u,\,v+\Theta f)\,du\left(dv+\Theta\, f_{,i}\, dx^i+\Theta\, f_{,z}\, dz\right)+F(u,\,v+\Theta\, f,\,x,\,z)\,du^2\\
\nonumber
&&+g\left(u,\,v+\Theta\, f\right)\,h_{ij}\left(x\right)\,dx^idx^j+dz^2\mbox{ ,}
\end{eqnarray}
and
\begin{eqnarray}
\nonumber
\tilde T&=&2\,\tilde T_{uv}\left(u,\,v+\Theta\, f,\,x,\,z\right)\,du\,\left(dv+\Theta\, f_{,i}\,dx^i+\Theta\, f_{,z}\,dz\right)+ \tilde T_{uu}\left(u,\,v+\Theta\, f,\,x,\,z\right)\,du^2\\
\nonumber
&+& \tilde T_{vv}\left(u,\,v+\Theta\, f,\,x,\,z\right)\,dv^2+ \tilde T_{ij}\left(u,\,v+\Theta\, f,\,x,\,z\right)\,dx^i\,dx^j+ \tilde T_{zz}\left(u,\,v+\Theta\, f,\,x,\,z\right)\,dz^2 \mbox{ ,}
\end{eqnarray}
where $\Theta=\Theta(u)$ is the Heaviside step function. Using the coordinate transformation $\hat u=u$, $\hat x=x$, $\hat z=z$ and $\hat v=v+\Theta\, f$ we get (after suppressing all hats)
\begin{equation}
ds_{n}^2=2\,A\left(u,\,v\right)\,dudv+F\left(u,\,x,\,z\right)\,du^2+g\left(u,\,v\right)\,h_{ij}\left(x\right)\,dx^idx^j+dz^2\mbox{ ,}
\end{equation}
where $F=-2\,A\left(u,\,v\right)f\left(x,\,z\right)\,\delta$ and $\delta\equiv\delta(u)$ is the Dirac-delta function. Now the energy momentum tensor reads
\begin{eqnarray}
\tilde T&=&2\left(\tilde T_{uv}- \tilde T_{vv}\,f\,\delta\right)\,du\,dv+\left(\tilde T_{uu}+ \tilde T_{vv}\,f^2\,\delta^2-2\, \tilde T_{uv}\,f\,\delta\right)\,du^2+ \nonumber\\&&+\tilde T_{vv}\,dv^2+\tilde T_{ij}\,dx^idx^j+ \tilde T_{zz}\,dz^2\mbox{ ,}
\end{eqnarray}
where the space we are interested in has $\tilde T^{\mbox{\scriptsize{background}}}_{vv}=\tilde T^{\mbox{\scriptsize{background}}}_{uu}=0$. 
 Plugging the above form of the energy momentum tensor into Einstein's equations $\tilde R_{\mu\nu}-\tilde g_{\mu\nu}\,\tilde R/2=\tilde T_{\mu\nu}/M_{n}^{n-2}$ we obtain~\cite{Sfetsos:1994xa} 
\begin{equation}
\label{einstein equation uu}
\tilde R_{uu}=\frac{1}{M_{n}^{n-2}}\tilde T_{uu}-2\,f\left(x,\,z\right)\delta\left(u\right)\tilde R_{uv} \mbox{ ,}
\end{equation} 
and
\begin{equation}
\label{u v einstein equation}
\tilde R_{uv}=\frac{1}{M_n^{n-2}}\left(\tilde T_{uv}-\frac{g_{uv}}{n-2}\tilde T \right)\mbox{ ,}
\end{equation}
where $M_{n}$ is the $n$-dimensional Planck mass.

The general strategy for obtaining the shock wave equation is to start by calculating the components of Ricci tensor for the $ds_n^2$ metric, then we add the contribution from the conformal factor~\footnote{Given the metric $d\tilde s^{\,2}_n=\Omega^2(z)ds_n^2$, the relation between $\tilde R_{\sigma\nu}$ and $R_{\sigma\nu}$ is given by $\tilde R_{\sigma\nu}=R_{\sigma\nu}-\left[(n-2)\delta_{\sigma}^{\alpha}\delta_{\nu}^{\beta}+g_{\sigma\nu}g^{\alpha\beta} \right]\Omega^{\,-1}\left(\nabla_\alpha\nabla_\beta\Omega\right)+\left[2(n-2)\delta_{\sigma}^{\alpha}\delta_{\nu}^{\beta}-(n-3)g_{\sigma\nu}g^{\alpha\beta} \right]\Omega^{-2}\left(\nabla_\alpha \Omega \right)\left(\nabla_{\beta}\Omega\right)$\,. }.
The final results are
\begin{eqnarray}
\nonumber
\tilde R_{uu}&=&\frac{n-3}{2}\left(\frac{g_{,u}A_{,u}}{gA}-\frac{g_{,uu}}{g}+\frac{g_{,u}^2}{2g^2}\right)+\left[\frac{2A_{,uv}}{A}+\frac{n-3}{2}\frac{g_{,uv}}{g}+\frac{A}{g}\Delta_{\,h}+A\partial^2_z+2A\frac{\partial^{2}_{z} \Omega}{\Omega}\right.\\
\nonumber
&+&\left.(n-2)A\frac{\partial_z\Omega}{\Omega}\partial_{z} +2A(n-3)\left(\frac{\partial z \Omega}{\Omega}\right)^2-\frac{2A_{,u}A_{,v}}{A^2}-\frac{n-3}{2g^2}g_{,u}g_{,v}\right.\\
\label{r uu}
&+&\left.\frac{n-3}{2Ag}\left( g_{,u}g_{,v}+g_{,v}A_{,u} \right)\right]f\delta+\left(\frac{A_{,vv}}{A}-\frac{A_{,v}^2}{A^2}+\frac{n-3}{2}\frac{g_{,v}A_{,v}}{gA}\right)f^2\delta^{2}\mbox{ ,}
\end{eqnarray} 
and
\begin{eqnarray}
\nonumber
\tilde R_{uv}&=&\left(\frac{A_{,u}A_{,v}}{A^2}-\frac{A_{,uv}}{A}+\frac{n-3}{4}\frac{g_{,u}g_{,v}}{g^2}-\frac{n-3}{2}\frac{g_{,uv}}{g}\right)+\left(\frac{A_{,v}^2}{A^2}-\frac{A_{,vv}}{A}-\frac{n-3}{2}\frac{g_{,v}A_{,v}}{gA} \right)f\delta\\
\label{ r uv} 
&-&\frac{\partial_{z}^2\Omega}{\Omega}A-(n-3)\left(\frac{\partial_{z}\Omega}{\Omega}\right)^2A\mbox{ ,}
\end{eqnarray}   
where $\Delta_h$ denotes the laplacian associated to the metric $h_{ij}$.

Now we consider $AdS_{n}$ space foliated into $AdS_{n-1}$ slices. In addition, we introduce a brane at $z=0$ with the $AdS_{n}$ space being $Z_2$ symmetric across the brane. The form of the metric is given by \cite{Kaloper:2005wq}
\begin{equation}
d\tilde s^{\,2}_{Ads_{n}}=\Omega^2(|z|)\left[\frac{4\,du\, dv}{\left(1-uv/\ell_{n-1}^{\,2}\right)^2}+\ell_{n-1}^{\,2}\,\left(\frac{1+uv/\ell_{n-1}^{\,2}}{1-uv/\ell_{n-1}^{\,2}}\right)^2\,\left(d\chi^2+\sinh^2\chi\, d\Omega_{n-4}^2\right) \right]\,\,.
\end{equation}
 where $d\Omega_{n-4}^2$ is the metric on the $(n-4)$-dimensional sphere, and the function $\Omega(|z|)$ is given by $\Omega\left(\left|z\right|\right)=\ell_n/\ell_{n-1}\,\sin\left((z_{0}+|z|)/\ell_{n-1}\right)$ where $\ell_{n-1}$ and $\ell_n$ are the radii of curvature of the brane and bulk space respectively. Direct calculations show that the first bracket in (\ref{r uu}) vanishes identically. In addition, $A_{,v}|_{u=0}=g_{,v}|_{u=0}=0$. The only nonvanishing component of the energy momentum tensor of a massless particle on the brane is given by
\begin{equation}
\tilde T_{uu}^{\mbox{\scriptsize{particle}}}=\frac{2p}{\ell_{n-1}^{n-3}\sqrt{h}}\,\delta(u)\,\delta(z)\,\delta(\chi)\,\delta(\theta_{1})\,\ldots \delta(\theta_{n-4})\mbox{ .}
\end{equation}
Using (\ref{einstein equation uu}) we get
\begin{equation}\label{eqfgen}
\partial_{z}^2 f+(n-2)\frac{\partial_{| z|}\Omega}{\Omega}\partial_{| z|}f+\frac{1}{\ell^2}\left(\Delta_{\,h}+3-n\right)f=\frac{p}{M_{n}^{n-2}}\frac{1}{\ell_{n-1}^{n-3}\sqrt h}\delta(z)\delta(\chi)\delta(\theta_{1})...\delta(\theta_{n-4})\mbox{ .}
\end{equation}

By solving this equation it is possible to find gravitational shock wave solutions associated to brane null sources in the Karch-Randall model of any dimensionality.

\smallskip

The above formulae allow us to find the relationship between bulk and brane cosmological constants, $\Lambda$ and $\sigma$ respectively, in terms of the bulk and brane radii. One can find such relation from the $u-v$ component of Einstein's equations (\ref{u v einstein equation}) using $T_{uv}=T_{uv}^{\mbox{\scriptsize{bulk}}}+T_{uv}^{\mbox{\scriptsize{brane}}}=-\Lambda g_{uv}-\sigma g_{uv}\delta(z)$ and $T^{\,\alpha}_{\alpha}=-n\Lambda-(n-1)\sigma\delta(z)$. Substituting (\ref{ r uv} ) into (\ref{u v einstein equation}) we get
\begin{eqnarray}
\nonumber
&-&A\left[\frac{-2\delta(z)}{\ell_{n-1}}\cot(z_{0}/\ell_{n-1}) +\frac{n-1}{\ell^2_{n-1}\sin^2\left((| z|+z_0)/\ell_{n-1}\right)} \right]\\
&=&\frac{1}{M_{n}^{n-2}}g_{uv}\left[-\Lambda\left(1-\frac{n}{n-2}\right)-\sigma\left(1-\frac{n-1}{n-2} \right)\delta(z)\right]\mbox{ .}
\end{eqnarray}
Noticing that $g_{uv}=A\,\Omega^2$ we find
\begin{equation}
\Lambda=-\frac{(n-1)(n-2)M_{n}^{n-2}}{2\ell_{n}^{\,2}}\mbox{ ,}
\end{equation}
and
\begin{eqnarray}
\sigma=2\,(n-2)\,M^{n-2}_n\sqrt{\frac{1}{\ell_{n}^{\,2}}-\frac{1}{\ell_{n-1}^{\,2}}}\mbox{.}
\end{eqnarray}

\smallskip

Finally, we can compute the CFT energy momentum tensor associated to a given shock wave solution. In order to do this, we compute the $n-1$ dimensional Einstein tensor built with the above metric restricted to the brane, and identify it with the CFT energy momentum tensor times $M_{n-1}^{n-3}$, where $M_{n-1}$ is the Planck mass on the brane. This way we obtain
\begin{equation}
\label{induced energy momentum tensor}
T_{uu}^{\mbox{\scriptsize {CFT}}}=\frac{2\,\left(n-3\right)\,M_{n-1}^{n-3}}{\ell_{n-1}^{\,2}}\left[\frac{\partial^2}{\partial\chi^2}-1 \right]f(\chi)\,\delta(u)\mbox{ .}
\end{equation}

\end{document}